\pdfoutput=1

\documentclass[11pt]{article}

\usepackage[]{coling}
\usepackage{times}
\usepackage{latexsym}
\usepackage[T1]{fontenc}
\usepackage[utf8]{inputenc}
\usepackage{microtype}
\usepackage{inconsolata}
\usepackage{graphicx}
\usepackage{xcolor}

\usepackage{hyperref}
\usepackage{color, colortbl}
\usepackage{booktabs}
\usepackage[algo2e,ruled,vlined]{algorithm2e} 
\usepackage{amsmath,amsfonts,bbm}
\usepackage{multirow}
\usepackage{booktabs,adjustbox}
\usepackage{svg} 
\definecolor{prompt}{RGB}{255,228,196}
\usepackage[symbol]{footmisc}

\usepackage[algo2e,ruled,vlined]{algorithm2e}

\title{TaCIE: Enhancing Instruction Comprehension in Large Language Models through Task-Centred Instruction Evolution}

\author{Jiuding Yang $^1$\ 
\textbf{Shengyao Lu} $^1$\
\stepcounter{footnote}Weidong Guo\thanks{\ \ Corresponding author.} $^2$\ 
\textbf{Xiangyang Li} $^2$\ \\
\textbf{Kaitong Yang} $^2$\ 
\textbf{Yu Xu} $^2$\ 
\textbf{Di Niu} $^1$\  \\
\text{$^1$University of Alberta}\\
\text{$^2$Platform and Content Group, Tencent}\\
\texttt{$^1$\{jiuding,shengyao,dniu\}@ualberta.ca}\\
\texttt{$^2$\{weidongguo,xiangyangli,kaitongyang,henrysxu\}@tencent.com}
}

\begin{document}
\maketitle
\begin{abstract}
Large Language Models (LLMs) require precise alignment with complex instructions to optimize their performance in real-world applications. As the demand for refined instruction tuning data increases, traditional methods that evolve simple seed instructions often struggle to effectively enhance complexity or manage difficulty scaling across various domains. Our innovative approach, Task-Centered Instruction Evolution (TaCIE), addresses these shortcomings by redefining instruction evolution from merely evolving seed instructions to a more dynamic and comprehensive combination of elements. TaCIE starts by deconstructing complex instructions into their fundamental components. It then generates and integrates new elements with the original ones, reassembling them into more sophisticated instructions that progressively increase in difficulty, diversity, and complexity. Applied across multiple domains, LLMs fine-tuned with these evolved instructions have substantially outperformed those tuned with conventional methods, marking a significant advancement in instruction-based model fine-tuning.
\end{abstract}

\section{Introduction}
\label{sec:intro}
The rapid development of Large Language Models (LLMs) and their expanding real-world applications necessitate closer alignment with complex human instructions to enhance performance across diverse tasks. This alignment demands high-quality instruction tuning data. However, manually crafting such instructions is impractical due to the time-intensive nature of the process and the tendency for these human-written instructions to remain simplistic  \citep{WizardLM}, offering minimal benefits for tuning effectiveness\cite{AMB}.

To mitigate the high costs and challenges of manual instruction creation, researchers have developed automated synthesis methods using powerful LLMs to generate more sophisticated instructions from simpler ones. Notable among these are \textsc{Self-Instruct} by \citet{self-instruct}, which expands the range of instructions from a set of seed inputs, and \textsc{Evol-Instruct} by \citet{WizardLM}, which refines instructions by enhancing either diversity or difficulty. \citet{IF} also introduced \textit{Instruction Fusion}, combining two distinct instructions to increase task complexity.

\begin{figure}[h]
\centering
\includegraphics[width=0.9\columnwidth]{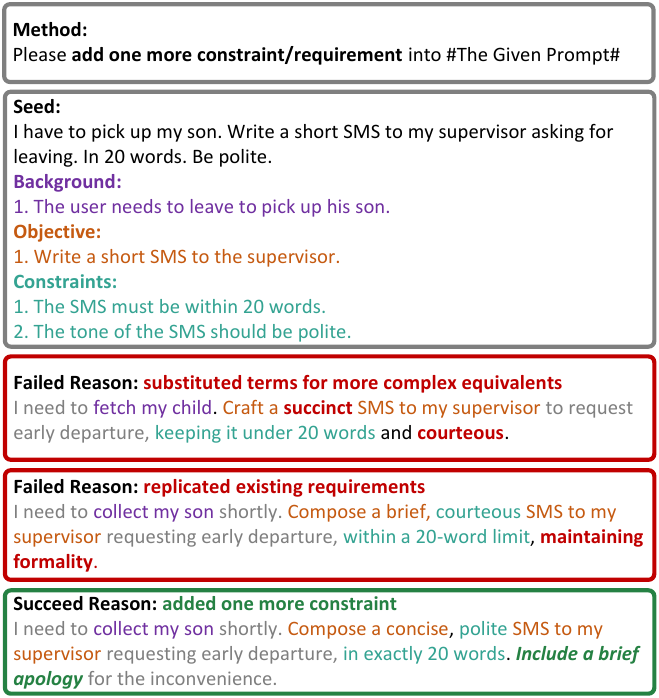} 
\caption{Real examples of applying \textsc{Evol-Instruct} using GPT-4.}
\label{fig-evol-example}
\vspace{-5mm}
\end{figure}

Despite their success in enhancing instruction quality and LLM performance, existing methods like \textsc{Evol-Instruct} and \textit{Instruction Fusion} exhibit significant limitations. Firstly, \textsc{Evol-Instruct} struggles with managing difficulty increments effectively; prompts such as ``add one more constrain'' often lead to vague enhancements that do not genuinely increase the task's difficulty. For example, a depth evolving experiment with GPT-4o\footnote{https://platform.openai.com/docs/models} showed that only one out of three attempts successfully intensified the instruction’s complexity (Figure~\ref{fig-evol-example}). Other attempts either replicated existing requirements or merely substituted terms for more complex equivalents, illustrating the challenges of uncontrolled difficulty scaling. Furthermore, \citet{WizardCoder}'s application in code generation tasks excessively escalated difficulty, adding seven constraints in just four rounds.

Secondly, these methods fail to adequately address cross-domain tasks. Although \textsc{Evol-Instruct} was implemented in both math and code generation tasks, it largely focused on increasing difficulty within the specific domain of the initial instruction, leading to a lack of diversity in task complexity. To mitigate this, \citep{IF} developed \textit{Instruction Fusion} to combine elements from two seed instructions. However, this approach, limited to a single round of fusion in code generation, falls short of fully exploiting the potential for enhanced complexity.

To address these deficiencies, we introduce Task-Centred Instruction Evolution (TaCIE), which employs advanced LLMs like GPT-4o to dissect and reassemble instructional elements, targeting enhancements in both difficulty and complexity. TaCIE systematically decomposes instructions into background information, objectives, and constraints. This allows for precise modifications and fosters more significant evolution of instructions. By shifting the focus from evolving individual instructions to evolving their basic elements, TaCIE not only refines difficulty scaling but also enables the integration of cross-domain elements. This approach significantly enhances the complexity and applicability of the evolved instructions.

Empirical results show that LLMs fine-tuned with TaCIE-evolved instructions outperform those tuned with existing methods on diverse benchmarks, including MT-Bench \citep{MT-Bench}, AlpacaEval \citep{AlpacaEval}, GSM8K \citep{GSM8K}, and HumanEval \citep{HumanEval}. This highlights TaCIE's ability to generate instructions that are more complex, nuanced, and broadly applicable across various domains. The key contributions of this work are:
\begin{itemize} 
\item We introduce TaCIE, a task-centered instruction evolution method, which decomposes seed instructions into three distinct elements and generates evolved instructions by modifying these elements. TaCIE effectively overcomes the challenges associated with difficulty scaling and cross-domain applicability, overcoming the limitation of existing instruction evolution methods such as \textsc{Evol-Instruct} and \textit{Instruction Fusion}.

\item Our extensive fine-tuning experiments across multiple domains demonstrate TaCIE’s superior performance in instruction generation. To promote research collaboration, we have open-sourced the model weights, training data, and source code, available at \textit{Anonymous}. \end{itemize}

\section{Approach}
\label{sec:method}
In this section, we propose TaCIE, a novel and efficient task-centred solution for instruction evolution to overcome the  limitations of existing methods. Before introducing TaCIE, we first discuss the existing instruction evolution method.

\subsection{Background}
\textsc{Self-Instruct} enhanced LLMs by fine-tuning with diverse self-generated instructions, leading to the development of \textsc{Evol-Instruct} by \citet{WizardLM}, which uses ChatGPT\footnote{https://chatgpt.com} to create more challenging and varied instructions from simple seed instructions. These seeds are evolved using five human-designed methods, significantly boosting LLM performance across various tasks. Recognizing limitations in task complexity, \citet{IF} introduced \textit{Instruction Fusion}, merging two seeds to enhance task complexity and performance, effectively complementing \textsc{Evol-Instruct}. Most recently, \citet{AIE} advanced this by developing an active instruction evolution method that uses GPT to select the optimal evolution strategy for seed instructions, further enhancing the instruction evolution's impact on LLM performance.

However, the methods discussed above share two significant limitations: inadequate management of difficulty increments and insufficient consideration of cross-domain tasks.

\textbf{Difficulty Increment Management} is essential for effective instruction evolution. Current methods struggle with this aspect due to vague prompts provided to LLMs, which lack specific guidance for evolving instructions. This results in uncontrollable and unpredictable outcomes. For instance, Figure~\ref{fig-evol-example} demonstrates the evolution of a simple instruction using \textsc{Evol-Instruct} and GPT-4o. Of three attempts, only the last one successfully added a new constraint to the seed instruction. The previous attempts either replaced terms with more complex equivalents or repeated existing requirements. For example, the second attempt evolved an instruction for a formal SMS, which GPT-4o interpreted in a manner too similar to the original, highlighting the inefficiency of these methods and their limited effectiveness in enhancing LLM fine-tuning.

\textbf{Cross-Domain Task Consideration} is equally critical. Despite improvements from methods like \textit{Instruction Fusion}, they fail to accommodate the complexity of cross-domain tasks. For instance, consider a task where an LLM is asked to develop an app function that automatically sends SMS messages in varying tones to selected contacts—this complexity far exceeds that of tasks focusing solely on message sending or SMS creation.

To address these challenges, we propose TaCIE, which is designed to effectively manage difficulty increments and cater to the demands of cross-domain tasks.

\begin{figure*}[h]
\centering
\includegraphics[width=0.9\textwidth]{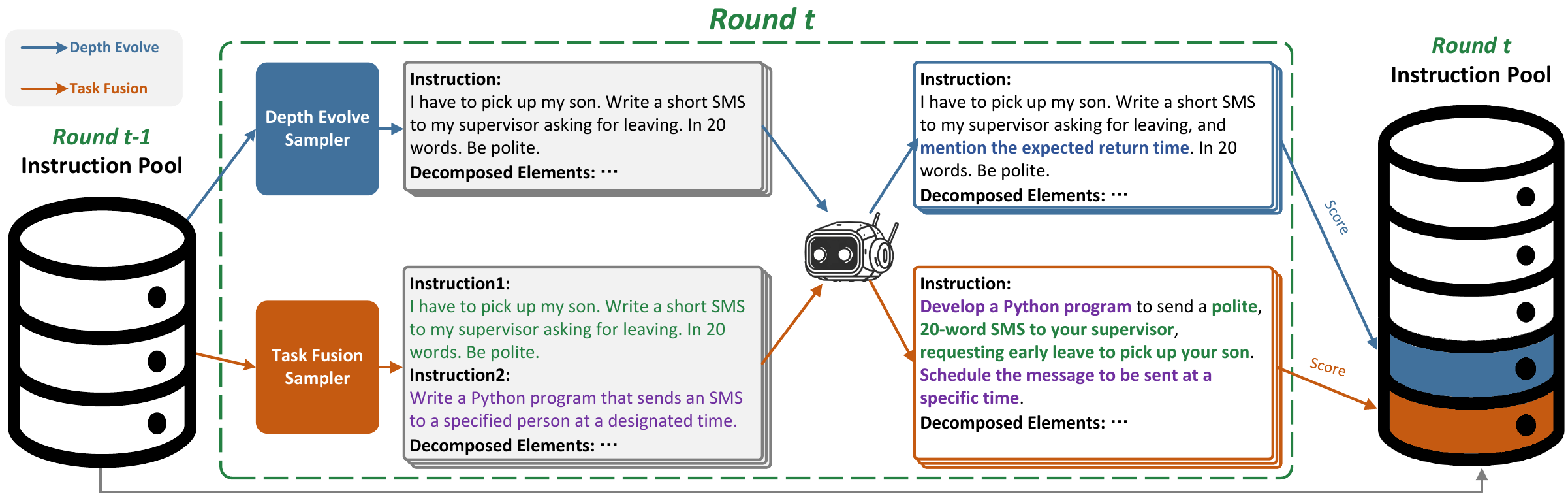} 
\caption{An illustration of TaCIE during the $t$-th round of evolution.}
\label{fig-illu}
\vspace{-5mm}
\end{figure*}

\subsection{Instruction Decomposition}
The most effective way to control incremental difficulty is to ensure that each new prompt introduces additional constraints or logical reasoning steps. To achieve this precise control over the evolution process, we draw inspiration from established decomposition methods in instructional design, as outlined in \citep{InfoBench, DeMoRecon}. We employ GPT-4o to dissect seed instructions into three fundamental components: Background, Objectives, and Constraints. This allows for direct modifications in constraint and logic reasoning applied to these elements. 

Figure~\ref{fig-evol-example} illustrates an example of this decomposition approach. The Background component encapsulates all relevant information necessary for the instruction, such as facts, motivations, and provided texts for tasks like summarization. The Objectives segment outlines the primary tasks derived from the seed instruction, for instance, composing an SMS as depicted in the example. Lastly, the Constraints section details specific requirements and limitations related to the tasks, including word count and formatting requirements.

Let \(\mathbf{C} = \{c_i\}_{1 \leq i \leq N}\) be the set of seed instructions, where \(c_i\) is the \(i\)-th instruction and \(N\) is the number of seeds. We define:
\begin{equation}
    \mathbf{E} = \{e_i\}_{1 \leq i \leq N} = \{\texttt{Decompose}(c_i)\}_{0 \leq i \leq N}.
\end{equation}
Here, \(e_i\) represents the decomposed elements of seed \(c_i\), \texttt{Decompose} denotes the decomposition process, and \(\mathbf{E}\) is the set of decomposed elements from the seed pool \(\mathbf{C}\). A detailed prompt template for decomposition is introduced in the Appendix.

\subsection{Task-Centred Instruction Evolution}
We break down the original seeds and then iteratively apply two types of evolution: depth evolution and task fusion, as illustrated in Figure~\ref{fig-illu}. 

For depth evolution, we aims to increase the difficult of newly generated instructions. To manage the increment in difficulty, we have developed a prompt template. This template guides the evolver to add precisely one additional constraint or one extra background setting to the elements of the seed instruction, thereby enhancing either the difficulty or the logical reasoning required. For example, as shown in Figure~\ref{fig-illu}, the depth evolution successfully added an extra constraint to the original instruction by requiring the mention of the expected return time in an SMS requesting leave. 

For task fusion, our objective is to enhance the complexity of tasks in fused instructions, making them more informative. We instruct the evolver to merge all elements from each pair of seed instructions, as illustrated in Figure~\ref{fig-illu}. The different colors in the orange box represent elements from the two seed instructions.

Despite the benefits these evolutionary methods offer to LLMs, selecting the right candidate instruction is crucial. A well-chosen seed can lead to evolved instructions of higher quality. To maximize the effectiveness of the proposed evolution methods, we have also developed a new candidate sampling technique. The entire process is segmented into the following stages:

\textbf{Seed Collection.}
We commence by collating seed instructions from a diverse assortment of open-source datasets tailored to various specializations: Alpaca \citep{alpaca} and ShareGPT \citep{sharegpt} for general instruction comprehension and execution; GSM8K \citep{GSM8K} and MATH \citep{MATH} for mathematical problem-solving; and CodeAlpaca \citep{WizardCoder} for coding tasks. Initially, we employ Sentence Transformers\footnote{https://huggingface.co/sentence-transformers/all-MiniLM-L6-v2} to compute embeddings for all instructions within these datasets. Using the Elbow method, we identify the optimal clustering configuration for each source. After clustering, we randomly select one seed instruction from each cluster. To broaden the diversity of our initial seed pool, we engage GPT-4o to generate three additional instructions that each have a different objective from the original seed. This approach enriches our collection of instructions, creating a robust foundational seed pool for the TaCIE framework, detailed further in the Appendix.

\textbf{Sample Scoring.}
The effectiveness of instruction tuning for LLMs relies on the quality of instruction-response pairs, rather than their quantity. As \citet{LIMA} have noted, instruction sets of higher quality confer more significant benefits to LLMs than those merely larger in volume. To curate such beneficial instructions, following \citet{AMB}, we employ uncertainty to filter out less beneficial instructions.

According to \citet{AMB}, it has been established that LLMs benefit more from fine-tuning with informative and novel instructions. Such instructions exhibit a significant alteration in the probability of their original responses when subjected to minor perturbations, like the random omission of a certain percentage of words. This alteration is considerably less pronounced in instructions that are less informative (low uncertainty), which are either ``easy'' (high response probability) or ``difficult'' (low response probability) to the LLMs. 

In the TaCIE framework, candidate sampling for the evolution process is guided by calculating uncertainty scores, which help identify instructions with potential for greater informativeness. The depth evolution strategy primarily targets seeds with high uncertainty, aiming to evolve these into more difficult instructions by incorporating additional logical reasoning steps or constraints. In contrast, task fusion merges information from two less informative seeds (low uncertainty) to create a single, more comprehensive instruction. This approach ensures that newly generated instructions are enriched with useful content without becoming overly complex, thereby enhancing their practical applicability in training LLMs. Further details on this process are elaborated in the Appendix.

Specifically, we define uncertainty $u_i$ as the score of the $i$-th instruction and use the following \texttt{Score} function to calculate it: 
\begin{equation}
    u_i = \texttt{Score}(c_i,r_i) = \frac{1}{N_U}\sum_{j=1}^{N_U}(|q_{i} - q_{i,j}|),
\end{equation}
where $N_U$ is the number of perturbation for each instruction, and
\begin{equation}
    q_{i,j} = \mathcal{P}(r_i|c_i,\mathbf{W}).
\end{equation}
Here, $r_i$ represents the response to the instruction $c_i$, and $q_{i,j}$ is the probability of response $r_i$ given instruction $c_i$ and the model weights $\mathbf{W}$. The index $j$ denotes the $j$-th perturbation of $c_i$. 

\textbf{Candidate Sampling.}
As mentioned by \citet{IF}, difficulty gradient is also an important key for better fine-tuning performance. To balance informative, ``easy'', and ``difficult'' instructions, we designed a sampler which samples the candidates for the depth evolution and task fusion according to different weighting approach. Denote the target evolution amount to be $M_e$ and $M_t$, for depth evolution, we directly use the uncertainty to weight each seed, and defined the sample probability of each instruction as:
\begin{equation}
    p^\texttt{depth}_{i} = \frac{u_i}{\sum_{k=1}^N u_k},
\end{equation}
and we sample $M_e$ instructions:
\begin{equation}
\mathbf{C}^\texttt{depth} \sim \texttt{Multinomial}(M_e, \{p^\texttt{depth}_{i}\}_{i=1}^{N}),
\end{equation}

For task fusion, we use the following weighting function:
\begin{equation}
    p^\texttt{fuse}_{i} = \frac{s_i}{\sum_{k=1}^N s_k}
\end{equation}
where
\begin{equation}
     s_i = \frac{1}{(n_{c_i}+1)\times n_{\texttt{obj}_i}\times n_{\texttt{root}_i}\times u_i}.
\end{equation}
Here $s_i$ is the punished uncertainty of the seed instruction $c_i$. For each instruction, we punish the uncertainty with three factors. $n_{c_i}$ represent the frequency of instruction $c_i$ being used as the seed for task fusion; $n_{\texttt{obj}_i}$ is the number of objectives of the instruction, which will increase if it is fused instruction last round; $n_{\texttt{root}_i}$ is the frequency of the root domain that instruction lies in, such as coding, math, etc. The sampling process is shown in the Algorithm~1.

\begin{algorithm2e}[h]\small
\caption{Task Fusion Sampling}\label{alg:tfs}
\SetAlgoLined
\textbf{Input:} Seed pool $\mathbf{C}$ which weights; Target number of task fusion $M_f$.\\
\textbf{Output:} Candidate pairs.

\textbf{Initialize} $\texttt{pairs}_{\text{in}}, \texttt{pairs}_{\text{cross}} \gets \emptyset$, $\texttt{idx} \gets 0$\\
\textbf{Let} $\mathbf{C}_a \sim \texttt{Multinomial}(M_f, \{p^\texttt{fuse}_{i}\}_{i=1}^{N})$\\
\While{$\texttt{pairs}_{\texttt{in}} < \frac{M_f}{2}$ \textbf{or} $\texttt{pairs}_{\texttt{cross}} < \frac{M_f}{2}$}{
    $\texttt{remain} \gets M_f - (|\texttt{pairs}_{\texttt{in}}| + |\texttt{pairs}_{\texttt{cross}}|)$\\
    $\mathbf{C}_b \sim \texttt{Multinomial}(\texttt{remain}, \{p^\texttt{fuse}_{i}\}_{i=1}^{N})$\\
    \For{\texttt{each} $c_b \in \mathbf{C}_b$}{
        $c_a \gets \mathbf{F}_a[\texttt{idx}]$\\
        \If{$\texttt{domain}(c_a) == \texttt{domain}(c_b)$}{
        \If{$|\texttt{pairs}_{\texttt{in}}|< \frac{M_f}{2}$}{
            $\texttt{pairs}_{\texttt{in}} \cup \{(c_a,c_b)\}$\\
            $\texttt{idx} \gets \texttt{idx} + 1$
        }}
        \Else{ \If{$|\texttt{pairs}_{\texttt{cross}}|< \frac{M_f}{2}$}{
            $\texttt{pairs}_{\texttt{cross}} \cup \{(c_a,c_b)\}$\\
            $\texttt{idx} \gets \texttt{idx} + 1$
            }
        }
    }
}
\Return{$\texttt{pairs}_{\texttt{in}} \cup \texttt{pairs}_{\texttt{cross}}$}
\end{algorithm2e}

We categorize task fusion into two types: in-domain fusion and cross-domain fusion. In-domain fusion integrates tasks within the same domain, whereas cross-domain fusion combines tasks from different domains to generate more complex outcomes. The process begins by sampling $M_f$ initial candidates, denoted as $C_a$, which form the first set of seeds. Subsequently, we sample another set of $M_f$ candidates, $C_b$, and pair each candidate with those in $C_a$ based on the differences in their domains. The sampling of $C_b$ and its pairing process are repeated until we achieve the desired number of pairs for both in-domain and cross-domain fusions.

For simplification, in a cross-domain fused instruction involving a pair $(c_a, c_b)$, we designate the root domain of the instruction as the root domain of $c_a$.

\textbf{Evolution with Evolver.} 
In the round $t$ of evolution, based on our designed weighting functions above, we first sample $M_d$ candidates for the depth evolution, and $M_f$ candidate pairs from the seed pool for the $i$-th round $\mathbf{C}^t$ for the task fusion, which are:
\begin{align*}
&\mathbf{C}^{t,\texttt{depth}}=\{c^{t,\texttt{depth}}_i\}_{1\le i \le M_d},\\
&\mathbf{C}^{t,\texttt{fuse}}=\{(c^{t,\texttt{fuse}}_{i,a},c^{t,\texttt{fuse}}_{i,b})\}_{1\le i \le M_f}.
\end{align*}
Next, we prompt evolver LLM to perform the two kinds of evolution and obtained the new instruction sets $\mathbf{D}^t=\{d^t_i\}_{1\le i \le M_d}$ and $\mathbf{F}^t=\{f^t_i\}_{1\le i \le M_f}$ correspondingly, where
\begin{align*}
    &d^t_i=\texttt{Evol}(c^{t,\texttt{depth}}_i,e^{t,\texttt{depth}}_i,\texttt{P}^\texttt{depth}),\\
    &f^t_i=\texttt{Evol}((c^{t,\texttt{fuse}}_{i,a},e^{t,\texttt{fuse}}_{i,a},c^{t,\texttt{fuse}}_{i,b},e^{t,\texttt{fuse}}_{i,b}),\texttt{P}^\texttt{fuse}).
\end{align*}
Here $\texttt{P}^\texttt{depth}$ and $\texttt{P}^\texttt{fuse}$ are the prompt template designed for the evolution, and it detailed in Appendix.

After that, we merge them in to the previous seed pool to form the new set of candidates $\mathbf{C}^{t+1}=\mathbf{C}^{t}+\mathbf{D}^t+\mathbf{F}^t$, and update the weight of all samples according to their scores and evolution history for the next round.

\subsection{Data Statistic}
\begin{figure}[t]
\centering
\includegraphics[width=0.7\columnwidth]{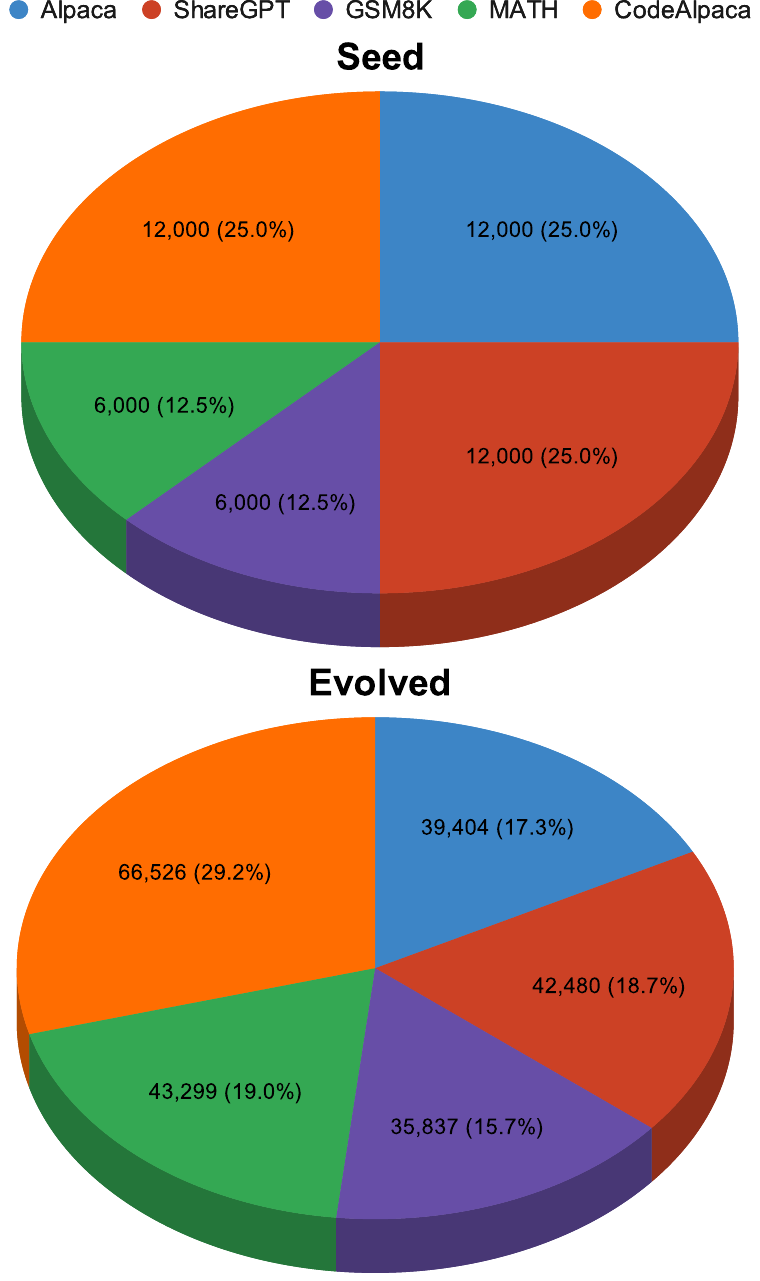} 
\caption{The domain distribution. Note each fused instruction contributes to multiple domains due to objectives from two seeds. }
\label{fig-statistic}
\vspace{-5mm}
\end{figure}
Figure~\ref{fig-statistic} presents detailed statistics from our evolved instruction pool. We sampled a total of 12,000 seeds from multiple sources: 3,000 each from ShareGPT, Alpaca, and Code Alpaca, along with 1,500 from both the MATH and GSM8K training sets. Using these seeds, we prompted GPT-4o to generate an additional 36,000 variants, aiming to diversify the seed pool with varied objectives. Over six evolutionary rounds—excluding samples that failed in evolution or were unrecognized by our scripts—we applied two distinct evolutionary methods, ultimately producing 143,917 viable samples out of 144,000 attempts. This yields a success rate of over 99.94\%, significantly surpassing the performance of \textsc{Evol-Instruct}. During the process, we utilized Llama-3-8B-Instruct\footnote{https://huggingface.co/meta-llama/Meta-Llama-3-8B-Instruct} to evaluate the uncertainty of each instruction.

Additionally, the figure illustrates frequency statistics across different domains. The scoring language model identified mathematics and coding problems as particularly "informative"—a finding consistent with expectations, given that these categories often require robust logical reasoning and exhibit considerable variability with minor instructional modifications. This shift in domain distribution has effectively improved the balance of our data mix, enhancing instruction tuning. Subsequent experiments confirm that this rebalancing has notably improved performance across all evaluated metrics.

\begin{figure}[h]
\centering
\includegraphics[width=0.9\columnwidth]{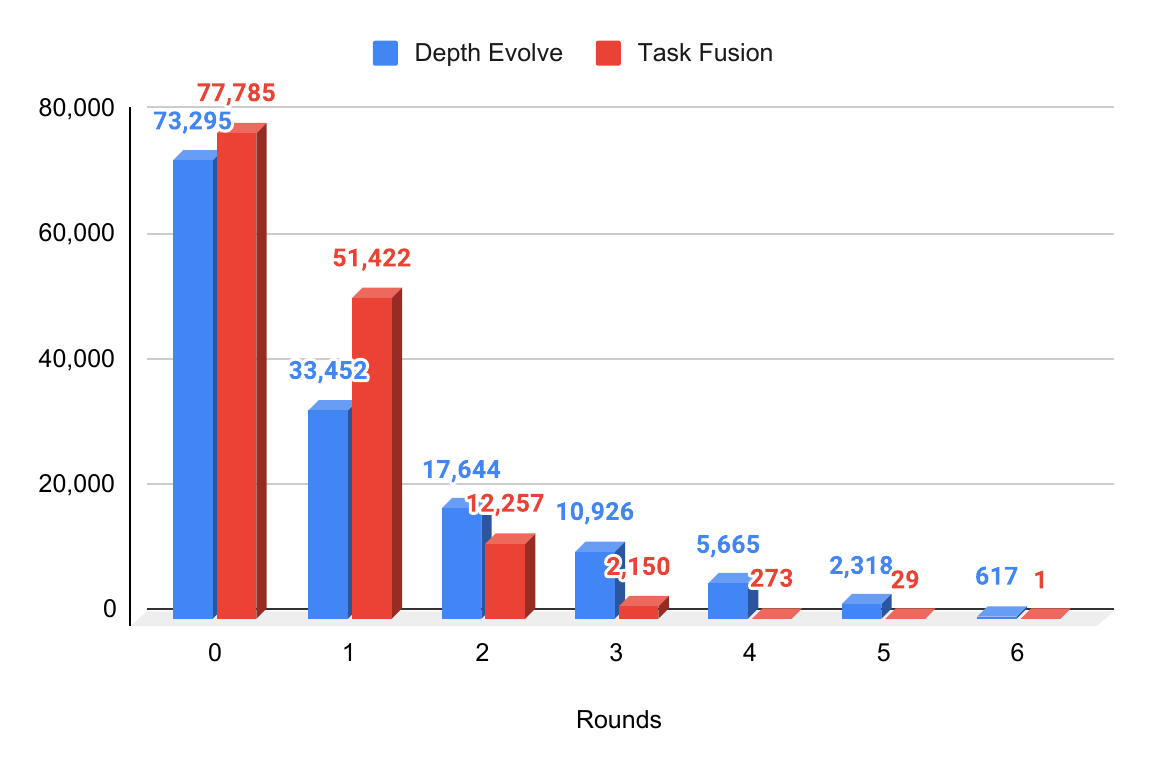} 
\caption{The distribution of evolution rounds.}
\label{fig-round}
\vspace{-5mm}
\end{figure}

Figure~\ref{fig-round} depicts the distribution of rounds for each evolutionary method. From this visualization, it is clear that TaCIE, in contrast to full-size evolution—which processes all candidates in each round and reaches 144,000 instructions in just three rounds—selects more informative seeds for evolution. This approach enables the generation of informative instructions over a higher number of evolutionary rounds within the same total number of generated instructions.

\section{Experiments}
\label{sec:exp}
\subsection{Experimental Settings}
To effectively showcase the capabilities of TaCIE, we selected a diverse range of baselines, encompassing both general-purpose chatting LLMs and domain-specific LLMs. Our primary comparisons are with \textsc{Evol-Instruct}, \textit{Auto Evol-Instruct}, and \textit{Instruction Fusion}. These three methods extensively utilize instruction evolution techniques, making them directly comparable to our approach. 

We assess performance across four dimensions. To evaluate general instruction comprehension, we utilize MT-Bench \citep{MT-Bench}. IFEval \citep{ifeval} is employed to test instruction-following capabilities. The GSM8K test set \citep{GSM8K} measures mathematical proficiency, and HumanEval \citep{HumanEval} gauges coding skills.

We utilized GPT-4o both as the evolver and the response generator for all instructions. Our foundational model for these experiments was LLaMA-3-8B \citep{llama3}, augmented with LLaMA-3-8B-Instruct for scoring purposes. Additionally, we fine-tuned Mistral-7B-v0.1 \citep{mistral} and Qwen2-7B \citep{qwen2} to evaluate the applicability of data generated with a dedicated scorer across different models. We also included performance metrics from their advanced official chat LLMs. However, it's important to note that their fine-tuning protocols often incorporate techniques beyond instruction tuning, such as Direct Preference Optimization \citep{DPO}. Therefore, their results are provided for reference only.

Consistent with existing literature, all our experiments were conducted using a batch size of 128 and a learning rate of \(5 \times 10^{-6}\). Each LLM was trained over four epochs using bfloat16 precision on four Nvidia A100 80G GPUs. For these processes, we utilized resources from LLaMA Factory \citep{llamafactory} and integrated DeepSpeed\footnote{https://github.com/microsoft/DeepSpeed} with zero-stage 2 optimization.

\subsection{Result Analysis}
\begin{table}[]
\begin{adjustbox}{max width=\columnwidth}
\begin{tabular}{lcccccc}\toprule
                                           Model& Para. & Data           & MT-Bench      & IFEval               & GSM8K          & HumanEval            \\ \midrule
\multicolumn{7}{c}{Close-Source Model}                                                                                                                  \\ \midrule
\multicolumn{1}{l|}{GPT-4}                 & -    & -                    & 8.99          & 85.37                 & 92             & 84.1                 \\
\multicolumn{1}{l|}{GPT-3.5}               & -    & -                    & 7.9           & -                     & 80.8           & 73.2                 \\ \midrule
\multicolumn{7}{c}{Open-Source General Model}                                                                                                           \\ \midrule
\multicolumn{1}{l|}{WizardLM}              & 13B  & 70k                  & 6.35          & 18.5                 & -              & 24                   \\ 
\multicolumn{1}{l|}{Vicuna-v1.3}           & 13B  & -                    & 6.57          & 33.44                & 10.77          & -                    \\
\multicolumn{1}{l|}{LLaMA-2-Chat}          & 13B  & -                    & 6.65          & 39.85                & 15.24          & 32.3                 \\
\multicolumn{1}{l|}{Mistral-instruct-v0.1} & 7B   & -                    & 6.84          & 42.51                & 14.25          & 31.1                 \\
\multicolumn{1}{l|}{LLaMA-3-Instruct}      & 8B   & -                    & 8.05          & 73.01                & 79.6           & 62.2                 \\
\multicolumn{1}{l|}{Qwen2-Instruct}        & 7B   & -                    & 8.41          & 55.08                & 82.3           & 79.9                 \\ \midrule
\multicolumn{7}{c}{TaCIE General Model}                                                                                                                 \\ \midrule
\multicolumn{1}{l|}{LLaMA-3-seed}          & 8B   & 48k                  & 7.14          & 38.45                & 64.82          & 43.9                 \\
\multicolumn{1}{l|}{TaCIE-LLaMA-3}         & 8B   & 48k                  & \textbf{7.18} & 39.74                & 68             & 51.2                 \\
\multicolumn{1}{l|}{TaCIE-LLaMA-3}         & 8B   & 144k                 & 6.63          & \textbf{42.7}        & \textbf{72.25} & \textbf{54.9}        \\\hline
\multicolumn{1}{l|}{Mistral-seed}          & 7B   & 48k                  & 6.75          & 27.73                & 53.3           & 21.3                 \\
\multicolumn{1}{l|}{TaCIE-Mistral}         & 7B   & 48k                  & \textbf{6.99} & 18.11                & 58.38          & 36.6                 \\
\multicolumn{1}{l|}{TaCIE-Mistral}         & 7B   & 144k                 & 6.73          & \textbf{30.13}       & \textbf{66.64} & \textbf{43.3}        \\\hline
\multicolumn{1}{l|}{Qwen2-seed}            & 7B   & 48k                  & 7.62          & 39                   & 82.56          & 72.6                 \\
\multicolumn{1}{l|}{TaCIE-Qwen2}           & 7B   & 48k                  & 7.85          & \textbf{43.99}       & \textbf{83.32} & 69.5                 \\
\multicolumn{1}{l|}{TaCIE-Qwen2}           & 7B   & 144k                 & \textbf{7.87} & 41.4                 & 81.65          & \textbf{76.8}        \\ \midrule
\multicolumn{1}{l|}{TaCIE-LLaMA-3-id}      & 8B   & 10k                  & 6.9           & 31.05          & 74             & \textbf{51.8} \\
\multicolumn{1}{l|}{TaCIE-LLaMA-3-cd}      & 8B   & 10k                  & \textbf{6.96} & \textbf{34.18} & \textbf{74.45} & \textbf{51.8} \\\midrule
\multicolumn{7}{c}{Open-Source Task-Specific}                                                                                                           \\ \midrule
\multicolumn{1}{l|}{AEI-ShareGPT}          & 7B   & 10k                  & 7.51          & -                    & -              & -                    \\
\multicolumn{1}{l|}{WizardMath}            & 7B   & 96k                  & -             & -                    & 54.9           & -                    \\
\multicolumn{1}{l|}{MetaMath}              & 7B   & 395k                 & -             & -                    & 66.51          & -                    \\
\multicolumn{1}{l|}{AEI-GSM8K}             & 7B   & 7k                   & -             & -                    & 70.74          & -                    \\
\multicolumn{1}{l|}{WizardCoder}           & 15B  & \multicolumn{1}{l}{} & -             & -                    & -              & \multicolumn{1}{l}{} \\
\multicolumn{1}{l|}{CodeLlama-Instruct}    & 13B  & -                    & -             & -                    & -              & 42.7                 \\
\multicolumn{1}{l|}{AEI-CodeAlpaca}        & 13B  & 20k                  & -             & -                    & -              & 65.85                \\
\multicolumn{1}{l|}{IF-20k}                & 13B  & 20k                  & -             & -                    & -              & \textbf{67.7}        \\ \midrule
\multicolumn{7}{c}{TaCIE Task-Specific}                                                                                                                 \\ \midrule
\multicolumn{1}{l|}{TaCIE-ShareGPT}        & 7B   & 10k                  & \textbf{7.53} & -                    & -              & -                    \\
\multicolumn{1}{l|}{TaCIE-GSM8K}           & 7B   & 7k                   & -             & -                    & \textbf{73.46} & -                    \\
\multicolumn{1}{l|}{TaCIE-CodeAlpaca}      & 13B  & 20k                  & -             & -                    & -              & 67.1    \\\bottomrule         
\end{tabular}
\end{adjustbox}
\caption{Experimental results of TaCIE.}
\label{tab-main}
\vspace{-5mm}
\end{table}
Table~\ref{tab-main} shows TaCIE's impact on LLaMA-3 across four benchmarks. We compared the performance of LLMs fine-tuned on 48,000 instructions from both evolved and seed datasets. The results demonstrate substantial improvements in instruction following, math, and coding tasks with the evolved instructions, while maintaining competitive performance on MT-bench. Using LLaMA-3-Instruct as the scorer, TaCIE effectively identifies and samples informative candidates, significantly enhancing task performance in these domains. Despite only a minor performance increment on MT-bench due to its multi-round chatting requirements—a feature not covered in our datasets—the overall gains from the evolved instructions outweigh this limitation. The domain shift focuses on complex tasks over multi-round chatting (Section 2.4), leading to less performance boost but considerable benefits across other domains, validating the evolution approach.

To justify the transferability of the evolved instructions (candidates sampled with the LLaMA-3 based scorer can also benefits other base LLMs), we use them to further fine-tune Mistral-7B-v0.1 and Qwen2-7B, which also show significant improvements all most domains. Here the beneftis for Qwen2 model is less than that on the other two base LLMs, because Qwen2-7B have better performance than LLaMA-3-8B\footnote{https://qwenlm.github.io/blog/qwen2/}, so what LLaMA-3-8B find informative may not be true for Qwen2-7B. However, it still outperforms the baseline fine-tuned with seeds.

Besides the general purpose LLMs, we also conduct experiment fine-tuning base models using domain-specific instructions. We mainly compare our performance with AEI proposed by \citet{AIE}. For fair comparison, we use the same base LLMs and sample the same amount of evolved instructions that only contains single-domain information among all of its evolution history. For ShareGPT, we mix 7,000 evolved instructions with 3,000 original samples to cover multi-round requirements of MT-Bench.

As shown in the domain-specific part in Table~\ref{tab-main}, we outperform AEI on all three domains, especially on math and coding, which requires better logic for answering. For \textit{Instruction Fusion}, we sampled 20,000 samples from their 110,000 evolved coding instructions for baseline fine-tuning (IF-20k). Although they has evolved more complex instructions with higher diversity due to the large amount of instructions, we still achieved 67.1\% on HumanEval, which is comparable to their 67.7\%.

In conclusion, TaCIE's evolved instructions significantly boost base LLM performance across tasks, particularly in complex areas like instruction comprehension, math, and coding. These enhancements persist even when applied to LLMs with varying architectures, such as Mistral-7B-v0.1 and Qwen2-7B, demonstrating their broad applicability and transferability. These findings confirm the effectiveness of the TaCIE evolution approach in enhancing instruction sets for diverse LLM applications.

\subsection{Ablation Study}
To further demonstrate task fusion's effectiveness, we fine-tuned LLaMA-3-8B with 10,000 single-domain and 10,000 cross-domain-only evolved instructions (excluding seeds). The results, shown as TaCIE-LLaMA-3-id and TaCIE-LLaMA-3-cd in Table~\ref{tab-main}, reveal that models fine-tuned with cross-domain instructions outperform those fine-tuned with in-domain instructions. This supports the notion that fusing objectives from different domains introduces greater complexity and information, enhancing LLM learning.

\begin{figure}[t]
\centering
\includegraphics[width=0.9\columnwidth]{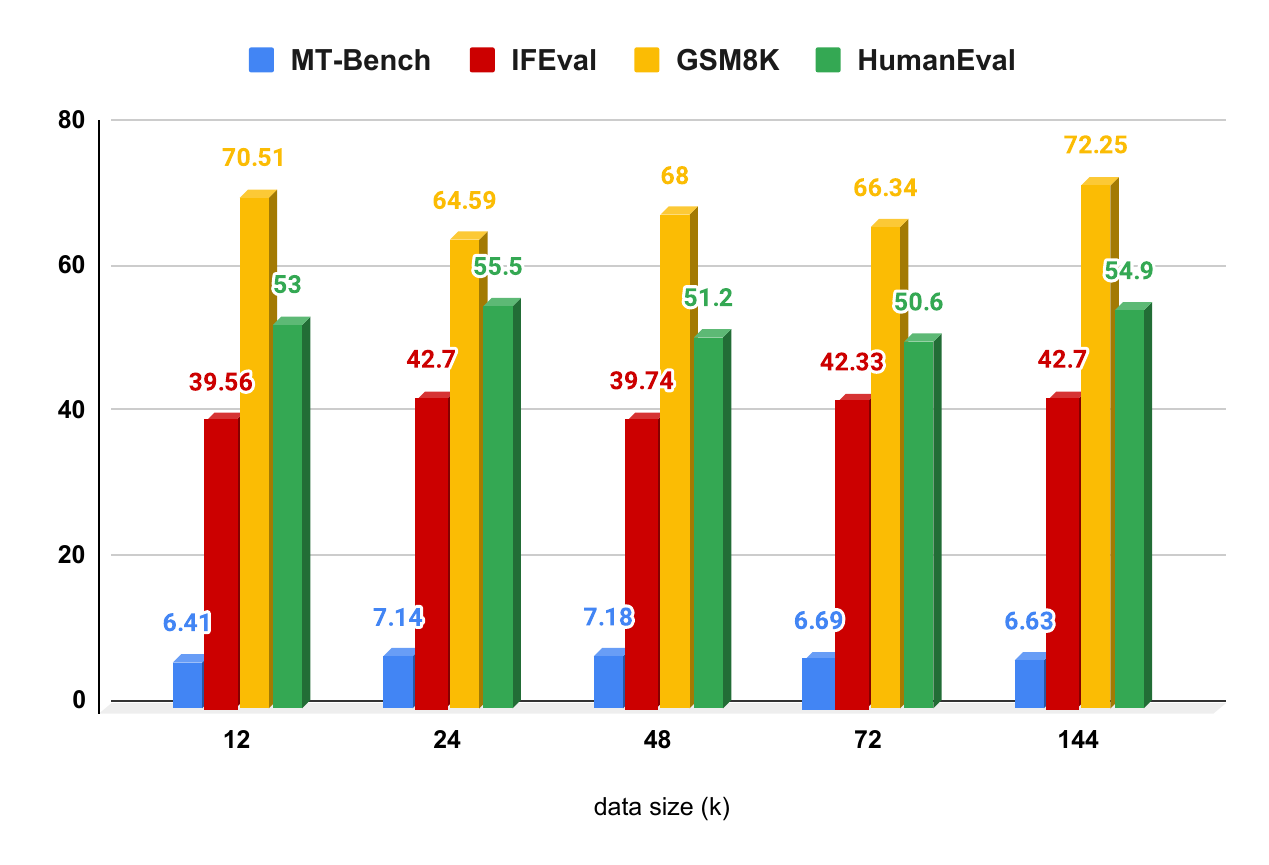} 
\caption{The performance scaling of TaCIE.}
\label{fig-scale}
\vspace{-5mm}
\end{figure}

Additionally, to evaluate the scalability of TaCIE, we fine-tuned LLaMA-3-8B with various subsets of instructions randomly sampled from our instruction pool. The results, depicted in Figure~\ref{fig-scale}, indicate that performance peaks when utilizing the entire instruction pool. This observation suggests that further enhancements in performance could be achieved through additional rounds of instruction pool evolution. We also extended this fine-tuning process to two other base LLMs using the complete set of 143,917 instructions (Table~\ref{tab-main}). This approach demonstrated significant improvements across multiple metrics when compared to using only 48,000 instructions.

\section{Related Work}
\label{sec:related}
\subsection{Instruction Tuning}
Instruction Tuning \citep{instruction-tuning} is a pivotal method for aligning Large Language Models (LLMs) with human instructions, enhancing their applicability to real-world scenarios \citep{it-survey1}. This approach aims to improve the zero-shot capabilities of well-trained LLMs, enabling them to perform tasks guided solely by natural human instructions. It allows LLMs to cater to a diverse array of general requests \citep{it-survey2}. However, the effectiveness of Instruction Tuning heavily depends on the availability of high-quality data to optimize both performance and interaction quality \citep{LIMA, long}.

For enhancing and proofing high-quality instructions, \citet{LIMA} introduced LIMA, an LLM trained on just 1,000 high-quality instances. This model demonstrated that smaller datasets of superior quality can be more beneficial than larger, less curated ones. Additionally, \citet{long} found that detailed, lengthy instructional responses also enhance LLM performance, underscoring the significance of data quality.

Conversely, \citet{AMB} explored the concept of task uncertainty by examining how minor perturbations in instructions affect response probabilities. They proposed that instructions whose perturbations can cause significant shifts in response probabilities are particularly informative and novel for LLMs, thus aiding in better alignment. This approach has inspired our methodology for the design of seed sampling.

These studies collectively highlight the critical role of high-quality instructions and responses in the efficacy of LLMs. Nonetheless, the reliance on manually crafted instructions or templates constrains the diversity, quantity, and creativity of the data available.

\subsection{Instruction Evolution}
In response to the growing demand for high-quality data, \citet{self-instruct} developed \textsc{Self-Instruct}, a strategy that utilizes LLMs for both generating data and tuning instructions. This method produces enhanced synthetic instructions and responses. Building on this concept, \citet{WizardLM} introduced \textsc{Evol-Instruct}, which evolves initial simple instructions into more challenging or varied forms by leveraging sophisticated LLMs (e.g., ChatGPT) and a designed evolution methodology. Further expanding on this, \citet{AIE} proposed an advanced framework that allows LLMs to autonomously determine the most effective evolution strategy for a given set of seed instructions.

\citet{IF} focused on task complexity by developing \textit{Instruction Fusion}, which integrates two simple tasks into a single, more complex challenge, thereby enhancing performance. This method has inspired further exploration into cross-domain task fusion within our proposed approach.

\section{Conclusion}
\label{sec:conclude}
In this paper, we introduce TaCIE, a novel method for evolving instruction that enhances difficulty management and promotes cross-domain complexity. TaCIE employs a decomposition approach to break down seed instructions into three fundamental elements. This transformation shifts the evolution process to the elemental level, allowing for targeted modifications that culminate in the regeneration of advanced instructions. Experimental results underscore TaCIE's efficacy, demonstrating significant performance improvements across a variety of base LLMs compared to previous methods.
\section*{Limitations}
Our work focus on task-centered evolution, making them more difficult and complex in controllable manner, thus the multi-turn chatting is not considered in our evolving procedure. Future work could easily built evolution method for multi-turn conversation based on our proposed method.

The total cost of our experimental process was approximately 2,000 USD, largely driven by our reliance on GPT-4o for augmentation and evaluation—a common challenge in LLM-related research. However, the expenses associated with using such APIs are decreasing due to rapid advancements in LLM technologies, making these tools more affordable and accessible.
\section*{Ethics Statement}
Our data collection relies on publicly released datasets. The augmented data were generated using GPT-4o, whose outputs are already carefully monitored, ensuring that no privacy-sensitive or confidential information was included.

\bibliography{custom}

\appendix
\section{Data Analysis}
\subsection{Diversification}
\begin{figure}[h]
\centering
\includegraphics[width=1\columnwidth]{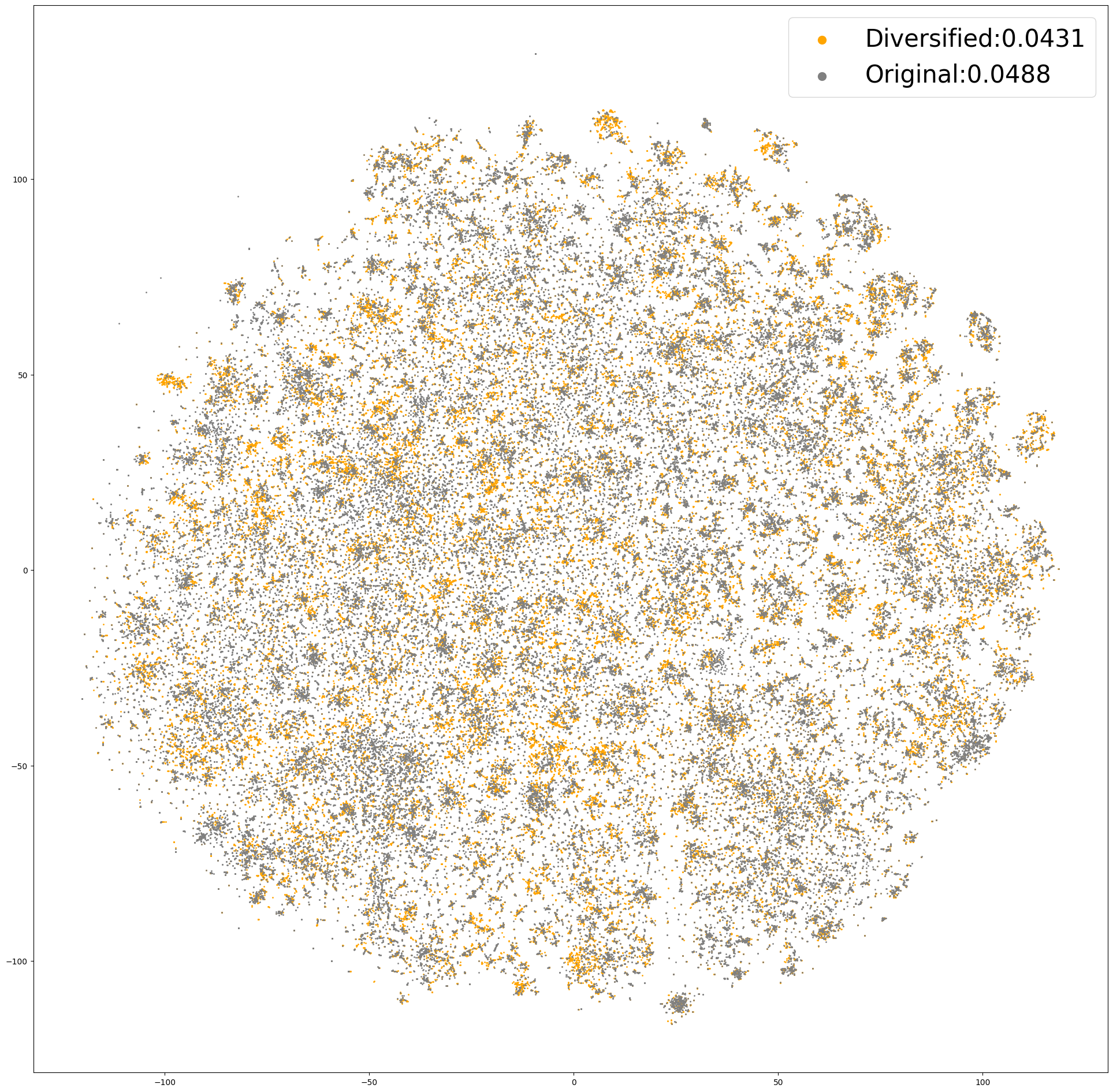} 
\caption{The 2D projection of original data source and diversified seed pool.}
\label{fig-diversity}
\end{figure}
As detailed in Section 2, we utilize the Elbow method to optimally cluster each data source and randomly select one instruction from these clusters as the initial seeds. These seeds are then used to prompt GPT-4o to generate a diverse set of 48,000 instructions, forming our initial instruction pool (Round 0). Figure~\ref{fig-diversity} shows a 2D projection of both the randomly sampled instructions from the original data sources and our diversified seed pool. It is evident that the diversified seeds cover more of the previously blank regions compared to the original data.

\begin{figure*}[t]
    \centering
    \begin{minipage}{0.32\textwidth}
        \centering
        \includegraphics[width=\linewidth]{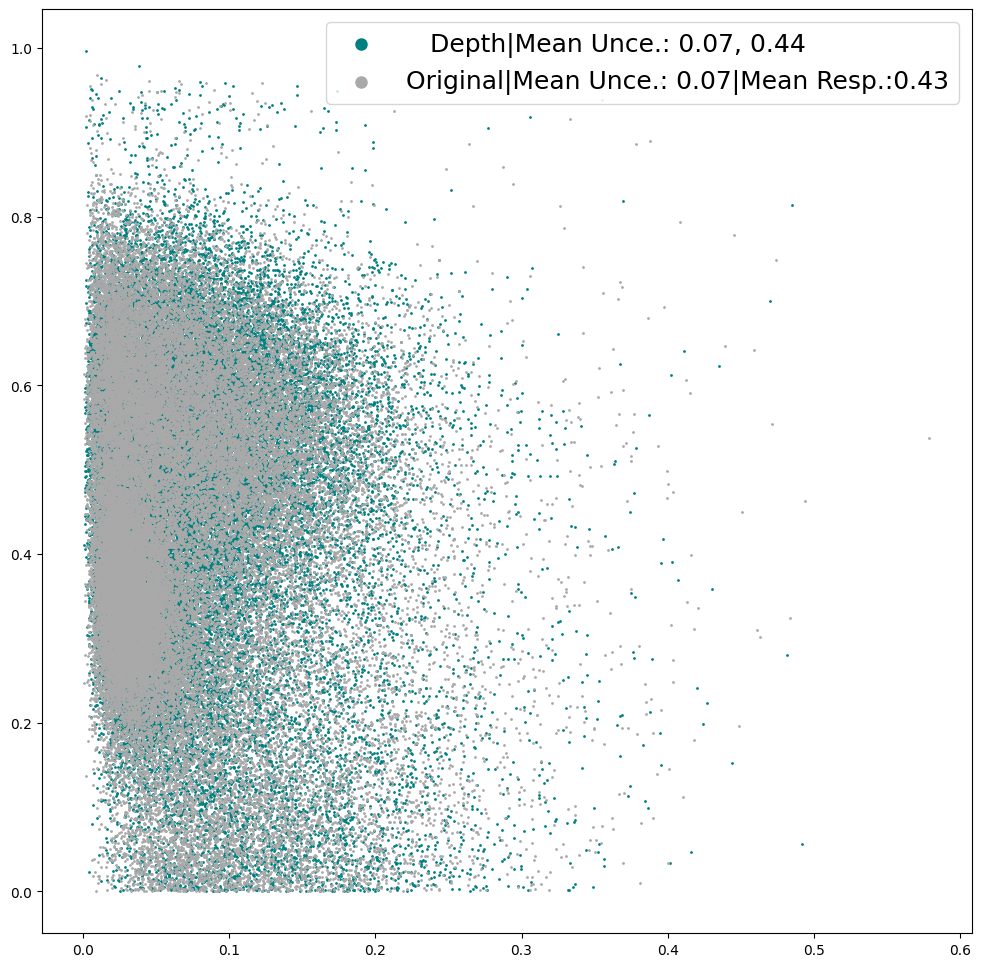} 
        \caption{Depth Evolution}
        \label{fig:shift1}
    \end{minipage}
    \hfill
    \begin{minipage}{0.32\textwidth}
        \centering
        \includegraphics[width=\linewidth]{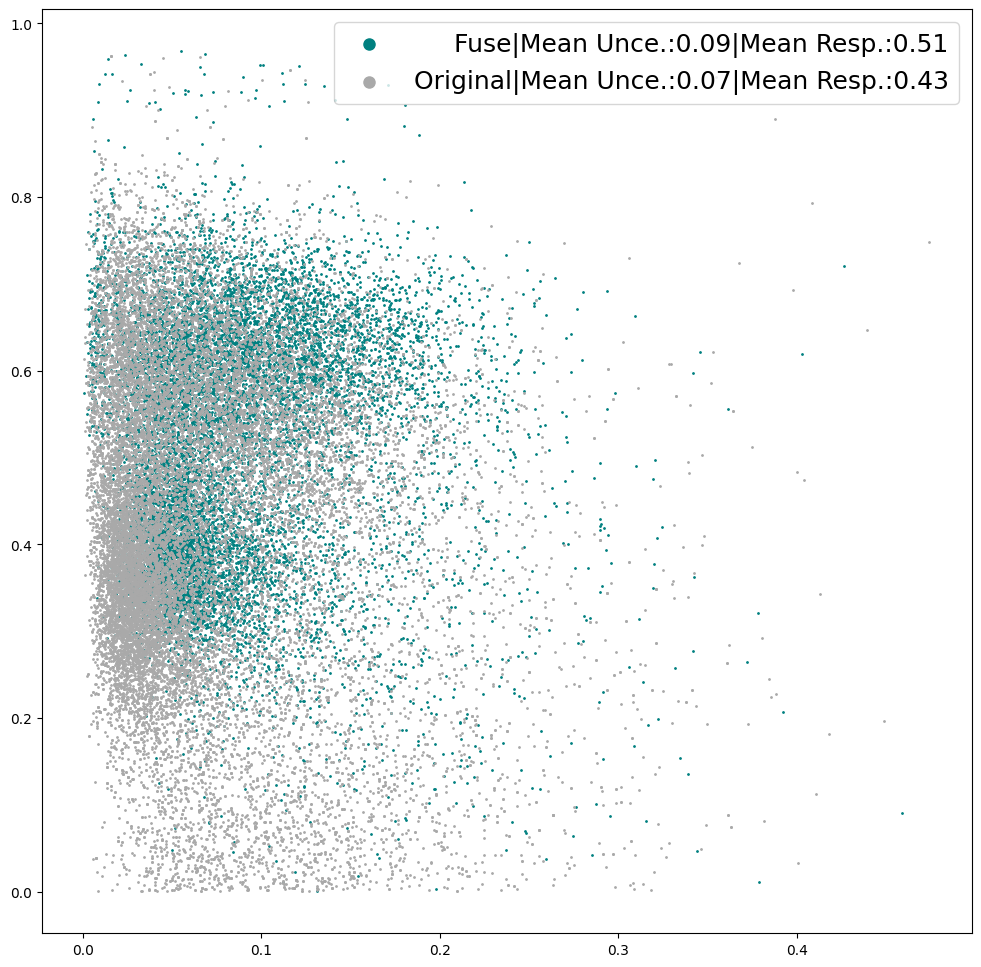} 
        \caption{In-domain Task Fusion}
        \label{fig:shift2}
    \end{minipage}
    \hfill
    \begin{minipage}{0.32\textwidth}
        \centering
        \includegraphics[width=\linewidth]{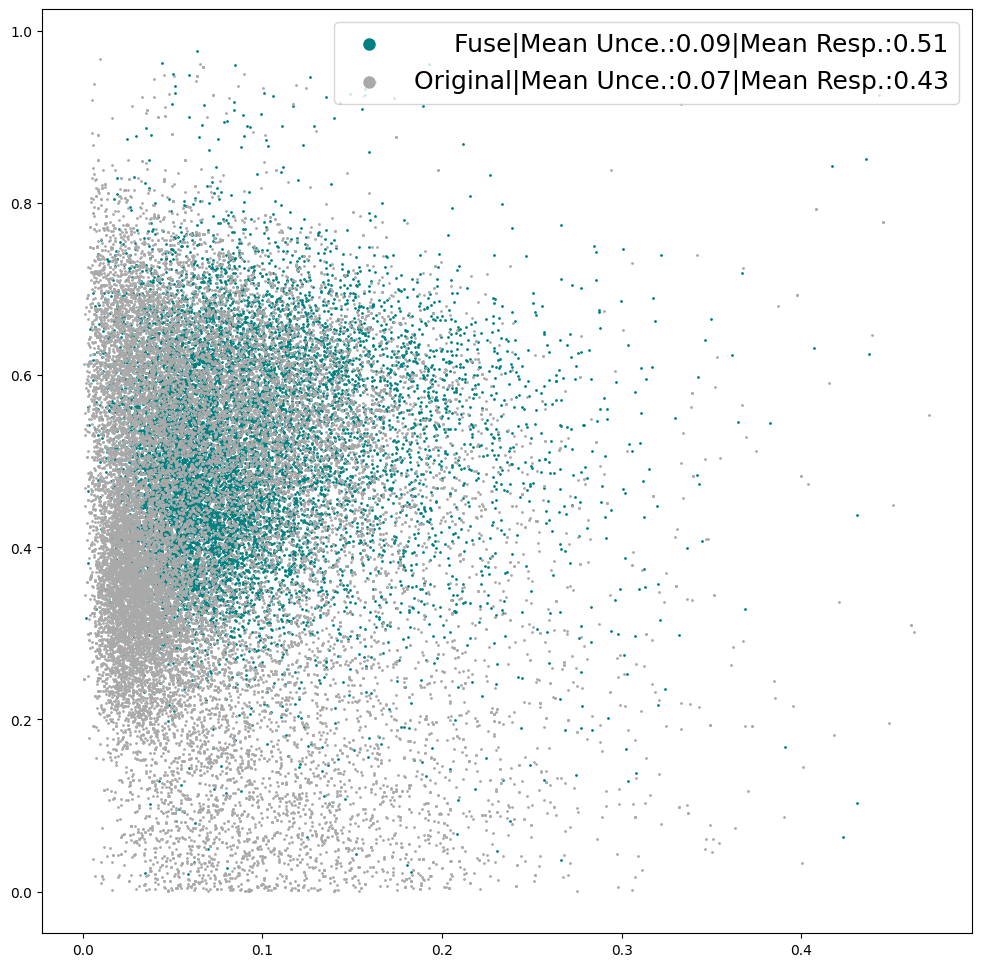} 
        \caption{Cross-domain Task Fusion}
        \label{fig:shift3}
    \end{minipage}
\end{figure*}

To quantify this increase in diversity, we employ the variance calculation method used in \textit{Instruction Fusion} by \citet{IF}:
\begin{equation}
U=\frac{1}{N}\Sigma^N_{i=1}(d_i-\mu)^2,
\end{equation}
where
\begin{equation}
d_i=||(e_i,e^\text{NN}_i)||
\end{equation}
and
\begin{equation}
\mu=\frac{1}{N}\Sigma^N_{i=1}d_i.
\end{equation}

In this formula, \(U\) represents the uniformity of the distribution, \(d_i\) denotes the Euclidean distance between the semantic embedding \(e_i\) of a seed and the embedding \(e^\text{NN}_i\) of its nearest neighbor, and \(\mu\) is the average Euclidean distance across all seeds. This calculation of variance in nearest neighbor distances provides a measure of the instruction pool's diversity. A lower variance signifies a more uniform distribution of data points, indicative of greater diversity. As depicted in the accompanying figure, the variance for the diversified seed pool stands at 0.0431, which is approximately 12\% lower than that of the original data source, confirming an enhancement in diversity.

\subsection{Uncertainty Shift}
To examine the impact of depth evolution and task fusion on instruction uncertainty, we generated 24,000 depth-evolved instructions and 12,000 task-fused instructions, both in-domain and cross-domain. Figure~\ref{fig:shift1}-\ref{fig:shift3} show 2D projections of the original and evolved instructions. From these figures, it is apparent that depth evolution does not significantly alter the uncertainty of instructions. In contrast, both in-domain and cross-domain task fusion lead to an increase in uncertainty, from 0.07 to 0.09. This indicates that task fusion, through the merging of two seed instructions, produces more informative content, thereby enhancing the overall uncertainty within the dataset. However, merging highly informative instructions could potentially overload the learning process of LLMs due to increased complexity. To mitigate this, during the sampling process, we prioritize less informative seeds, balancing the generation of enriched instructions while ensuring they remain tractable for learning enhancements.

\section{Prompt Templates}
In this section, we introduce all prompt templates we designed and use in TaCIE.

\subsection{Instruction Decomposition}
To better decompose each instruction, we provided GPT-4o with two examples during the decomposition process.

\noindent\emph{Given a prompt, your task is to:\\
1.**Extract Backgrounds Settings:** Identify and list the backgrounds of the prompt, such as facts, and motivations. If the prompt provides extra information such as code to debug, passage to polish or summarize, directly include them in this section, do not summarize them. Do not include Objectives or Requirements. If no Backgrounds Settings is given in the prompt, output 'N/A' in the **Extract Background Settings:** section.\\
2.**Extract Objective:** List the core task of the prompt.\\
3.**Extract Constraints:** List all specific requirements or constraints for the objectives. If no Constraints is given in the prompt, output 'N/A' in the **Extract Constraints:** section.\\
\\
---\\
\\
**Given Prompt:**\\
I have to pick up my son. Write a short SMS to my supervisor asking for leaving. In 20 words. Be polite.
\\
**Extract Background Settings:**\\
1.The user needs to pick up his son.\\
\\
**Extract Objectives:**\\
1.Write an SMS.\\
\\
**Extract Constraints:**\\
1.The SMS should be short.\\
2.The SMS should ask the supervisor's permission to leave.\\
3.The SMS should be polite.\\
4.The SMS should not exceed 20 words.\\
\\
---\\
\\
**Given Prompt:**\\
Debug my python code. a=100\\
b=1\\
c=0\\
print(d=a*b/c)\\
\\
**Extract Background Settings:**\\
1.Code to debug:\\
\textasciigrave\textasciigrave\textasciigrave python\\
a=100\\
b=1\\
c=0\\
print(d=a*b/c)\\
\textasciigrave\textasciigrave\textasciigrave\\
\\
**Extract Objectives:**\\
1.Debug the given Python code.\\
\\
**Extract Constraints:**\\
N/A\\
\\
---\\
\\
**Given Prompt**:\\
At 30, Anika is 4/3 the age of Maddie. What would be their average age in 15 years?\\
\\
**Extract Background Settings:**\\
1.age of Anika = 30\\
2.age of Anika = 4/3 x age of Maddie\\
\\
**Extract Objectives:**\\
1.Calculate the average age of Anika and Maddie in 15 years.\\
\\
**Extract Constraints:**\\
N/A\\
\\
---\\
\\
**Given Prompt:**\\
\{\{  prompt  \}\}\\
}

\subsection{Diversification}
During the diversification of the original seed, we use the following prompt to let GPT-4o generate another three different instructions for a given seed.

\noindent\emph{**For the provided prompt, we detail the following elements:**\\
\\
1.**Background Settings:** This section presents the background information relevant to the prompt, including pertinent facts and motivations. If the prompt lacks background settings, this section will be labeled as 'N/A'.\\
2.**Objectives:** This section outlines the main tasks associated with the prompt.\\
3.**Constraints:** This lists any specific requirements or limitations tied to the objectives. If there are no constraints, this section will be labeled as 'N/A'.\\
\\
\text**Given Prompt:**\\
\{\{  prompt  \}\}\\
\\
\{\{  extracted  \}\}\\
\\
Based on the information provided, your task is to:\\
\\
1.Develop ten new objectives to replace the original **Objectives** of the prompt. Each new objective should be diverse and maintain the same level of difficulty as the original.\\
2.For each new objective, craft a corresponding prompt that mimics the tone and style of the original prompt. Ensure to vary the **Background Settings** and **Constraints** while making sure each prompt is reasonable and answerable.\\
\\
Format each new objective and prompt as follows,do no provide corresponding background, objectives, and constraints:\\
\\
**New Objective 1:**\\
\text{[Describe the new diverse objective.]}\\
\\
**New Prompt 1:**
\text{[Present the new prompt based on the objective.]}\\
...}

\subsection{Depth Evolution}
\noindent\emph{Based on a prompt's existing background, objectives, and constraints, increase its difficulty using ONLY one of the following methods:
1.If the prompt primarily involves reasoning, such as solving a mathematical problem, enhance its complexity by introducing an additional background element. Also, modify the existing background elements to ensure the task remains logical and solvable.\\
2.Otherwise, introduce one additional reasonable constraint to ONLY one of the objectives of the given prompt to increase its difficulty.\\
\\
Please respond using the format provided in the examples below. You can only change either the **Background Settings:** or the **Constraints:**. Do not change both.\\
\\
Ensure your response contains the following four sections even if they are empty: **Prompt:**, **Background Settings:**, **Objective:** and **Constraints:**\\
---\\
\\
\#\#\# Original\\
\\
**Prompt:**\\
At 30, Anika is 4/3 the age of Maddie. What would be their average age in 15 years?\\
\\
**Background Settings:**\\
1.age of Anika = 30\\
2.age of Anika = 4/3 x age of Maddie\\
\\
**Objectives:**\\
1.Calculate the average age of Anika and Maddie after 15 years\\
\\
**Constraints:**\\
N/A\\
\\
\#\#\# Rewritten\\
\\
**Prompt:**\\
At 30, Anika's age is twice the age of Adam, and Maddie's age is the average of Anika's and Adam's ages. What would be the average age of Anika and Maddie in 15 years?\\
\\
**Background Settings:**\\
1.age of Anika = 30\\
2.age of Anika = 2 x age of Adam\\
3.age of Maddie =  (age of Anika + age of Adam) / 2\\
\\
**Objectives:**\\
1.Calculate the average age of Anika and Maddie after 15 years\\
\\
**Constraints:**\\
N/A\\
\\
---\\
\\
\#\#\# Original\\
\\
**Prompt:**\\
Could you write me an android application that has a login page and can connect to a server? Please also list all prior knowledge I need to know to understand your code.\\
\\
**Background Settings:**\\
N/A\\
\\
**Objectives:**\\
1.Write an Android application.\\
2.List all prior knowledge that is required to understand the code.\\
\\
**Constraints:**\\
1.Include a login page in the application.\\
2.Enable the application to connect to a server.\\
\\
\#\#\# Rewritten\\
\\
**Prompt:**\\
Could you write me an android application that has a login page, can connect to a server, and encrypts all communications with the server. Please also list all prior knowledge I need to know to understand your code.\\
\\
**Background Settings:**\\
N/A\\
\\
**Constraints:**\\
1.Include a login page in the application.\\
2.Enable the application to connect to a server.\\
3.Encrypt all communications with the server.\\
\\
---\\
\\
\#\#\# Original\\
\\
**Prompt:**\\
\{\{  prompt  \}\}\\
\\
\{\{  extracted  \}\}\\
}

\subsection{Task Fusion}
\noindent\emph{Based on the prompt's existing background, objectives, and constraints, your task is to act as a Prompt Fusion Specialist. Your target is to fuse **Given Prompt A** and **Given Prompt B** into a single, cohesive **Fused Prompt**, following the two steps below:\\
1.Merge the elements in the background, objectives, and constraints of both **Given Prompt A** and **Given Prompt B** respectively, make sure the objectives are dependent to each other and solvable. Do not compress multiple elements into a single one.\\
2.Based on the integrated background, objectives, and constraints, fuse the **Given Prompt A** and **Given Prompt B** into a new prompt. Mimic the tone and style of the original prompts. Make sure the new prompt is coherent and solvable\\
\\
---\\
\\
**Example Given Prompt A:**\\
I have to pick up my son. Write a short SMS to my supervisor asking for leaving. In 20 words. Be polite.\\
\\
**Background Settings:**\\
1.The user needs to pick up his son.\\
\\
**Objectives:**\\
1.Write an SMS.\\
\\
**Constrains:**\\
1.The SMS should be short.\\
2.The SMS should ask the supervisor's permission to leave.\\
3.The SMS should be polite.\\
4.The SMS should not exceed 20 words.\\
\\
**Example Given Prompt B:**\\
I am planning to give you a voice, and communicate through the speech medium. I need a speech recognizer, a wake call detector, and a speech synthesizer for your voice. Suggest a python script utilizing existing libraries to achieves the goal.\\
\\
**Background Settings:**\\
1.The user is planning to give a voice to a system and communicate through speech.\\
2.The user needs a speech recognizer, a wake call detector, and a speech synthesizer.\\
\\
**Objectives:**\\
1.Suggest a Python script using existing libraries to achieve the goal.\\
\\
**Constraints:**\\
N/A\\
\\
**Fused Background Settings:**\\
1.The user needs to pick up his son.\\
2.The user is planning to give a voice to a system and communicate through speech.\\
3.The system requires a speech recognizer, a wake call detector, and a speech synthesizer.\\
\\
**Fused Objectives:**\\
1.Suggest a Python script using existing libraries that enables a system to recognize speech, detect wake calls, and synthesize speech.\\
2.Use this system to compose and send an SMS.\\
\\
**Fused Constraints:**\\
1.The SMS should be short.\\
2.The SMS should ask the supervisor's permission to leave.\\
3.The SMS should be polite.\\
4.The SMS should not exceed 20 words.\\
5. The SMS should be composed by the system and sent to the user’s supervisor.\\
\\
**Fused Prompt:**\\
Suggest a Python script utilizing existing libraries that includes a speech recognizer, a wake call detector, and a speech synthesizer to enable communication through speech for a system I am planning to implement. Use this system to send a short, polite SMS to my supervisor asking for permission to leave early because I have to pick up my son. The message should not exceed 20 words. Be Polite.\\
\\
---\\
\\
Please respond using the format provided in the example above. Give the merged background, objectives, and constraints respectively, then fuse the two given prompts into a new one incorporating your new background, objectives, and constraints.\\
\\
**Given Prompt A:**\\
\{\{  prompt1  \}\}\\
\\
\{\{  extracted1  \}\}\\
\\
**Given Prompt B:**\\
\{\{  prompt2  \}\}\\
\\
\{\{  extracted2  \}\}\\}


\end{document}